# Neutron spectroscopic factors of Ni isotopes from transfer reactions

Jenny Lee (李曉菁), M.B. Tsang (曾敏兒), W.G. Lynch (連致標)

National Superconducting Cyclotron Laboratory and Department of Physics and Astronomy, Michigan State University, East Lansing, Michigan 48824, USA

M. Horoi

Department of Physics, Central Michigan University, Mount Pleasant, Michigan 48859, USA

S.C. Su (蘇士俊)

Physics Department, Chinese University of Hong Kong, Shatin, Hong Kong, China

## Abstract

177 neutron spectroscopic factors for nickel isotopes have been extracted by performing a systematic analysis of the angular distributions measured from (d,p) transfer reactions. A subset of the extracted spectroscopic factors are compared to predictions of large-basis shell models in the full *pf* model space using the GXPF1A effective interaction, and the $(f_{5/2}, p_{3/2}, p_{1/2}, g_{9/2})$ model space using the JJ4PNA interaction. For ground states, the predicted spectroscopic factors using the GXPF1A effective interaction in the full *pf* model space agree very well with the experimental values, while predictions based on several other effective interactions and model spaces are about 30% higher than the experimental values. For low-energy excited states (<3.5 MeV), the agreement between the extracted spectroscopic factors and shell model calculations is not better than a factor of two.

The shell structure of the unstable doubly magic nucleus $^{56}$Ni (N=Z=28) has attracted much attention recently [1-5]. In most shell model calculations, the N=28 core in $^{48}$Ca is assumed to be a well-established closed shell. However, Relativistic Hartree+Bogoliubov calculations predict a strong suppression of the N=28 shell gap for neutron rich nuclei [6]. While experimental investigations of the $2^+$ energies of $^{36,38,40}$Si provide evidence for the weakening of the N=28 shell gap in nuclei with large neutron excess [7], the evidence is inconclusive for the case of $^{47}$Ar [8,9]. Recent measurements of the nuclear magnetic moment of the ground state of $^{57}$Cu, which could be viewed as a valence proton outside a closed $^{56}$Ni core, suggests significant breaking of the $f_{7/2}$ shell [4]. To further explore the property of the single particle states outside $^{56}$Ni, we extracted the neutron spectroscopic factors, which measure the occupancy of the valence neutrons, ranging from $^{57}$Ni to $^{65}$Ni isotopes. The extracted spectroscopic factors are important bench marks in evaluating different *pf*-shell model interactions that may be used to predict the structure of Ni or Cu nuclei, particularly the doubly-magic nucleus $^{78}$Ni which is a major waiting point in the path of the r-process [10]. All these are of special importance for the stellar evolution and electron capture in supernovae.

Recent analysis [11] shows that aside from some states near the closed shell $^{40}$Ca nucleus, the experimental neutron spectroscopic factors obtained from excited states and large-basis shell-model predictions agree to better than 40% for Z=3-24 nuclei (Figure 3 of ref. [11]). However, there are large discrepancies between the excited state neutron spectroscopic factors and predictions for the Ni nuclei. Since only a subset of the SF's which have matched states in the shell model calculations are included in the global analysis [11], the main objective of the current paper is to publish all the neutron spectroscopic factors obtained from the analysis of the Ni isotopes. The present work also presents comparison of the ground state neutron spectroscopic factors in details. Unlike the excited states, the ground state neutron spectroscopic factors can distinguish different interactions used in large-basis shell model (LB-SM) calculations. We found that the data agree with the LB-SM predictions when the GXPF1A effective interaction in the full *pf* model space is used, while the agreement is not as good when other interactions are used.

For the present work, we extract the neutron spectroscopic factors of nucleus B in the reactions of A(d,p)B and B(p,d)A where the nucleus B is considered to be composed of the core A plus the valence neutron n. Following previous work [11-14] the experimental spectroscopic factor is defined as the ratio of measured cross sections to the cross sections calculated with a reaction model. A priori, transfer reactions do not yield absolute spectroscopic factors as the analysis depends on other input parameters such as the geometry of the neutron bound state wave function as well as the optical potentials used in the reaction model [14]. However, if the analysis utilizes a consistent set of parameters, the relative spectroscopic factors could be determined reliably [11-14]. We choose the analysis approach of references [12-14] as the ground state spectroscopic factors obtained agree with the large-basis shell model predictions to ~20%, which are similar to the experimental uncertainties of the extracted spectroscopic factors [12-14]. The reaction model used for the (p,d) and (d,p) reactions is the Johnson-Soper adiabatic three-body model that calculates the theoretical angular distributions [15] assuming unity spectroscopic factors. The Adiabatic Distorted Wave Approximation (ADWA) model takes into account the deuteron breakup in the mean field of the target. In the reaction model calculations, the global nucleon-nucleus optical potentials in ref. [16] are adopted. The potential binding the transferred neutron to the core is chosen to have Woods-Saxon shape with fixed radius and diffuseness parameters, $r_0$=1.25 fm and $a_0$=0.65 fm. The depth of the central potential well is adjusted to reproduce the experimental separation energies. All calculations make the local energy approximation (LEA) for finite range effects [17] using the zero-range strength ($D_o^2$=15006.25 MeV$^2$ fm$^3$) and range ($\beta$=0.7457 fm) parameters of the Reid soft-core $^3S_1$-$^3D_1$ neutron-proton interaction [18]. Nonlocality corrections with range parameters of 0.85 fm and 0.54 fm are included in the proton and deuteron channels [19]. We use the same source of input parameters for all the reactions analyzed here. The transfer reaction calculations are carried out using the code TWOFNR [20] which respects detailed balance between (p,d) and (d,p) reactions that connect the same states.

We extracted the neutron spectroscopic factors of $^{57,59,61,62,63,65}$Ni isotopes, using the angular distributions measured in (d,p) reactions found in the literature[21-35]. We supplement these data sets with neutron ground state spectroscopic factors determined from $^{58,60,61,62}$Ni(p,d) reactions [36-47] which should be the same as those determined from the inverse (d,p) reactions from detailed balance. To evaluate the effectiveness of the residual interactions and the model space used in the pf shell, these values will be compared to predictions from shell model calculations, SF(LB-SM), with four effective interactions and their associated model spaces.

From the published angular distributions [21-47], which are of reasonable quality, we extracted 177 spectroscopic factors, SF(ADWA), for the Ni isotopes. These values are listed in Table I. When available, spectroscopic factors from the Evaluated Nuclear Structure Data File (ENSDF) compiled by the National Nuclear Data Center (NNDC) [48] are also listed in Table I in the last column. In general, SF(ENSDF) values are taken directly from the published values, which came from different experiments and might be analyzed differently using different optical potentials and different reaction models. As a result, these SF values may not be consistent with each other or with the results from the present work. Fig. 1 compares the spectroscopic factors obtained in this work, SF(ADWA), (y-ordinate) to those listed in ENDSF (x-abscissa). The solid line indicates perfect agreement. Most of the ENDSF values are about 30% larger than the values obtained in the present work. (The spectroscopic factors for the data set $^{61}$Ni(d,p)$^{62}$Ni are not included in the comparison because of the discrepancies between the ENSDF and the values published in ref [30]. The second set of ENDSF values obtained from reference [49] for the same reaction does not have published angular distributions.)

Shell model calculations for Ni isotopes have been available since 1960's. In the early calculations [50-51], $^{56}$Ni is assumed to be an inert core and the influence of core excitation was taken into account in the effective residual interaction between the valence nucleons in the *pf*-shell. With advances in computational capability, many new effective interactions, which are the key elements for successful predictions, have been proposed. The GXPF1A interaction, a modified version of GXPF1 with five matrix elements, involving mostly the $p_{1/2}$ orbitals, has been obtained by adjusting the parameters used in the interaction to the experimental data [52]. Another interaction KB3 [53,54] has also been used to predict properties in the *pf* shell nuclei. The matching between the theoretical and experimental levels is based on the exact agreement of the quantum numbers ($l,j$) and spin-parity $J^{\pi}$ of the transferred neutron and the approximate agreement of the energy of the states. In general, the agreement between energy levels is within 300 keV. Both of these calculations require full *pf* model space and intensive CPU cycles. Recently, a new T=1 effective interaction for the $f_{5/2}$, $p_{3/2}$, $p_{1/2}$, $g_{9/2}$ model space has been obtained for the $^{56}$Ni-$^{78}$Ni region by fitting the experimental data of Ni isotopes from A=57 to A=78 and N=50 isotones for $^{89}$Cu to $^{100}$Sn [55]. This interaction provides an improved Hamiltonian for Z=28 with a large model space and new Hamiltonian for N=50. It has been mainly used to describe heavier Ni isotopes using a $^{56}$Ni core. Following the convention established in ref [56], this new interaction is called JJ4PNA in the present work. (The same interaction was called XT in ref. [11] and NR78 in ref. [57].) The predictions from the JJ4PNA interaction [55], using Oxbash code [58], as well as predictions from the GXPF1A interaction using the Antoine code [59] are listed in Table II.

Only limited numbers of states with excitation energy larger than 3 MeV have been calculated with the GXPF1A interaction because of the difficulties in identifying the states at high excitation energy and the CPU time required to do the calculations. Furthermore, all the states in *fp* shell have the same parity assignments ("-" for odd and "+" for even nuclei). On the other hand, more states are calculated and identified with the JJ4PNA interaction. In both calculations, the number of states, which have corresponding matched states in the shell model calculations, decreases with energy of the levels. For these reasons, a large number of experimental states has no counterparts in the shell model predictions and Table II lists 43 levels as compared to 177 experimental levels listed in Table I.

To demonstrate the sensitivity of the spectroscopic factors to interactions used in the shell model calculations, we first obtained the ground state neutron spectroscopic factors with the Auerbach interactions [50] and JJ4PNA interactions [55] using Oxbash code [58]. For calculations with GXPF1A and KB3 interactions we use the m-scheme code Antoine [59]. The latter calculations are CPU intensive. The comparison of the ground-state spectroscopic factors between experiments and calculations are shown in Fig. 2. The solid lines are the least square fits of the linear correlations between data and predictions. The slopes of the lines are labeled inside each panel. The predicted spectroscopic factors using the KB3, JJ4PNA, and Auerbach interactions are about 25% larger than the experimental values. The results using the full *pf* model space and the GXPF1A interaction, shown in the upper left panel of Fig. 2, give better agreement with the data as indicated by the slope (0.93±0.06) of the solid line in the upper left panel. This is consistent with the observation that with the improved modification in the monopole and pairing matrix elements of the Hamiltonian, the GXPF1A interaction is better than KB3 for the lighter isotopes around $^{56}$Ni [61,62]. This overall agreement with the results from GXPF1A interaction, is consistent with the trends established in nuclei with Z=3-24 [13]. Such agreement is in contrast to spectroscopic factors obtained from knockout reactions, which are quenched with respect to the shell model predictions depending on the neutron separation energy [60]. The same disagreement is also observed here. For the ground state of the $^{57}$Ni nucleus, the extracted spectroscopic factor from transfer reaction obtained in the present work is 0.95±0.29 but the spectroscopic factor from knockout reaction is 0.55±0.11 [2] while the best shell model prediction is 0.78 as discussed above. Currently there is no satisfactory explanation why the spectroscopic factors obtained in transfer reactions should be different from spectroscopic factors obtained from knockout reactions. Resolving such ambiguity may shed lights to the reaction mechanisms in transfer and knockout reactions.

Figure 3 shows the ratios of the experimental SF values to predicted SF values as a function of the energy levels for all the states we can identify in shell model calculations with GXPF1A interaction (top panel) and with JJ4PNA interaction (bottom panel). The solid lines (ratio=1) indicate perfect agreement between data and theory. The states, which are predicted by calculations using either the GXPF1A or JJ4PNA interactions but not both, are represented by the symbols with double edges. The current analysis yields spectroscopic factors that cluster around the large-basis shell model predictions. Based on experimental errors, the expected scattering of the data should be around 30%, within the dashed lines above and below the solid lines.

Below 3.5 MeV, predictions with the JJ4PNA interactions are in reasonable agreement with experimental data for light mass Ni isotopes with A~60 even though the interaction was developed to describe the heavy Ni isotopes around $^{67}$Ni. Aside from ground states (where the predictions by GXPF1A are better as shown in Fig. 2) and the light Ni isotopes (A<60), the scatter of the ratios in Fig.3 is similar in both calculations. Since the discrepancy between the data and the predictions significantly exceeds the experimental uncertainties shown by the error bars, the inaccuracies in the predictions mainly reflect the ambiguities in the interactions used in the calculations.

Above 3.5 MeV, there are only three states matched with predictions using the JJ4PNA interactions. The ratios of spectroscopic factors obtained for 4.709 MeV ($9/2^+$) and 5.429 MeV ($9/2^+$) states of the $^{59}$Ni nucleus, and 3.686 MeV ($3/2^-$) state in $^{61}$Ni are 7 and 3 and 9 respectively. The spectroscopic factors for all of them disagree with the shell model predictions beyond the systematics plotted in Figure 3. This suggests that properties of single particle energy levels at high excitation energy are not well described by the shell models even though the centroid of single particle energy may be determined from calculations especially when the hole states are taken into account [4].

More insights regarding the residual interactions may be obtained by combining the spectroscopic factors with energy level information. Each panel in Figure 4 compares the experimental energy levels and SF values to corresponding values obtained from shell model calculations for one isotope using the GXPF1A and JJ4PNA interaction. The lengths of the horizontal bars represent the values of the spectroscopic factors. As described earlier, very few states above 2 MeV have been obtained in the full *pf* model space using GXPF1A interaction. Figure 4 only show states with energy levels up to 2 MeV of $^{57,59,61,62,63,65}$Ni nuclei. In the upper left panel of Fig. 4, only three states have been measured for $^{57}$Ni. The description of the data by both calculations is quite reasonable. In $^{61}$Ni, the ordering of the states is not reproduced by calculations using any one of the two interactions. In general, shell model calculations tend to predict larger spectroscopic factors for the low-lying states, thus assigning larger single particle characteristics to these states. Due to the limitation of model space, no $g_{9/2}$ states (dashed lines in $^{63}$Ni and $^{65}$Ni nuclei) are predicted by calculations using the GXPF1A interactions.

In summary, neutron spectroscopic factors have been extracted for a range of Ni isotopes. The current set of measured spectroscopic factors provides an additional means other than energy levels to test the shell model interactions in the *pf* and $(f_{5/2}, p_{3/2}, p_{1/2}, g_{9/2})$ model spaces. For the ground state neutron spectroscopic factors, the calculations based on the GXPF1A effective interaction in the full *pf* model space give the best agreement with the data. For the excited states of Ni isotopes beyond $^{60}$Ni, the JJ4PNA effective interaction predicts the spectroscopic properties of these nuclei reasonably well. Agreement between data and shell model energy levels and spectroscopic factors deteriorates with excitation energy. For excited states below 3.5 MeV, the extracted spectroscopic factors cluster around the shell model values, but the agreement of the spectroscopic factors between data and calculations is not better than a factor of two. Since the experimental uncertainties are in the order of 20-30%, the data can be used to evaluate newer interactions in the *pf* and $(f_{5/2}, p_{3/2}, p_{1/2}, g_{9/2})$ model spaces. Improvement of interactions in the *pf* model spaces will be important to understand structural properties of the double magic nuclei of $^{78}$Ni.

**Acknowledgement**

The authors would like to thank Professors B.A. Brown and J. Tostevin for fruitful discussions. We also thank Dr. Lisetskiy for his help in the use of JJ4PNA interaction. This work is supported by the National Science Foundation under grants PHY-0606007, (JL, MBT, WGL) and PHY-0555366 (MH). Horoi acknowledges the NSF MRI grant PHY-0619407, which made possible the full *pf* model space shell model calculations with GXPF1A interaction. Su acknowledges the support of the Summer Undergraduate Research Experience (SURE) program sponsored by the Chinese University of Hong Kong.

Fig. 1. (Color online) Comparison of the extracted SF(ADWA) values in the present work and the compiled values in ENDSF [48] for both ground and excited states. The solid line indicates perfect agreement and dashed lines represent ±30% of solid line.

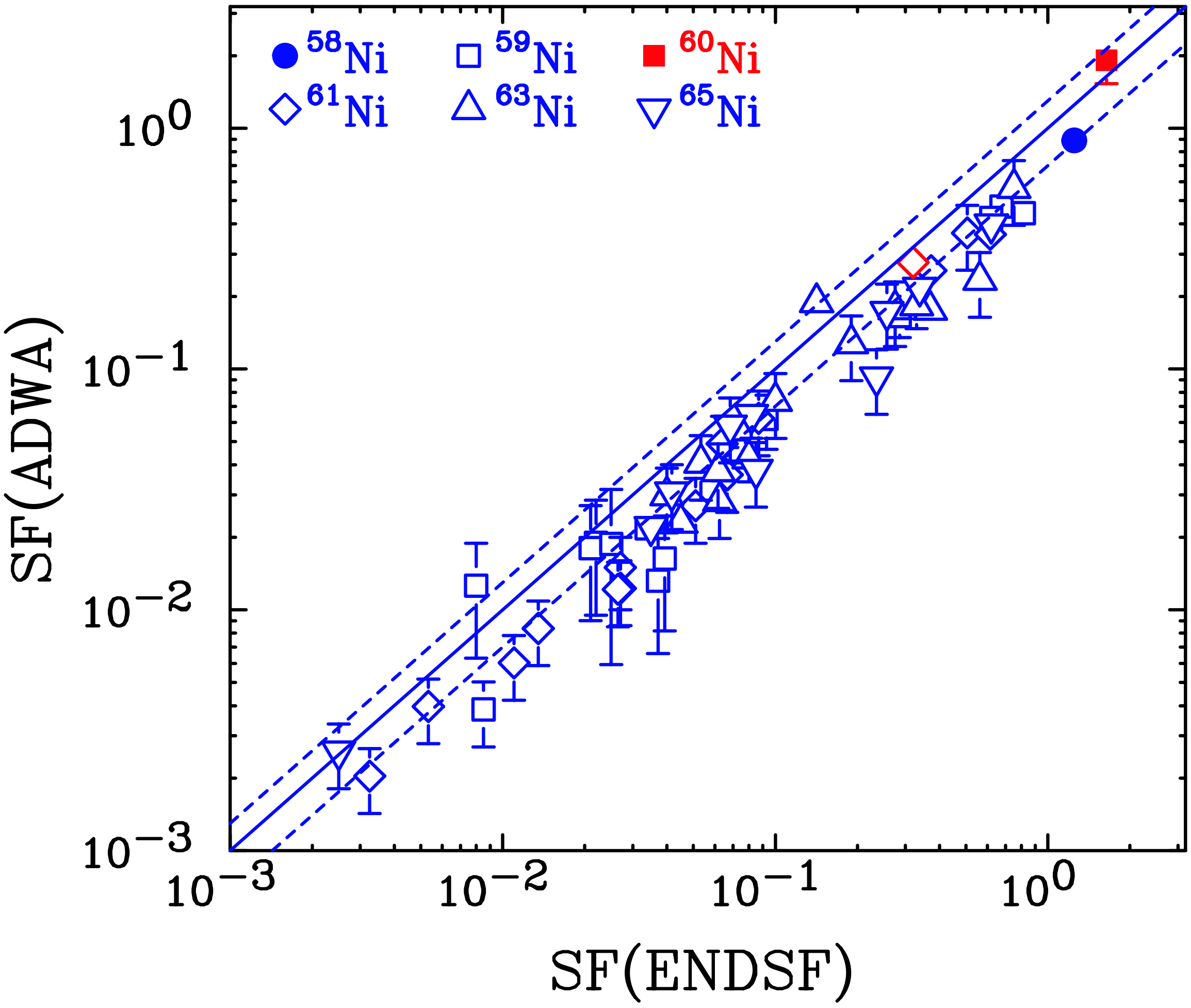

Fig. 2. (Color online) Left panels: Comparison of the ground-state experimental SF(ADWA) values and the shell model calculations with GXPF1A (top) and KB3 (bottom) interactions in full *pf* model space. Right panels: Same as left panels, but with calculations using JJ4PNA interaction in *gfp* model space (top) and calculations from N. Auerbach[50] (bottom). The solid lines are the linear fits with y-intercept fixed at zero. The slopes of the lines are listed in the individual panels.

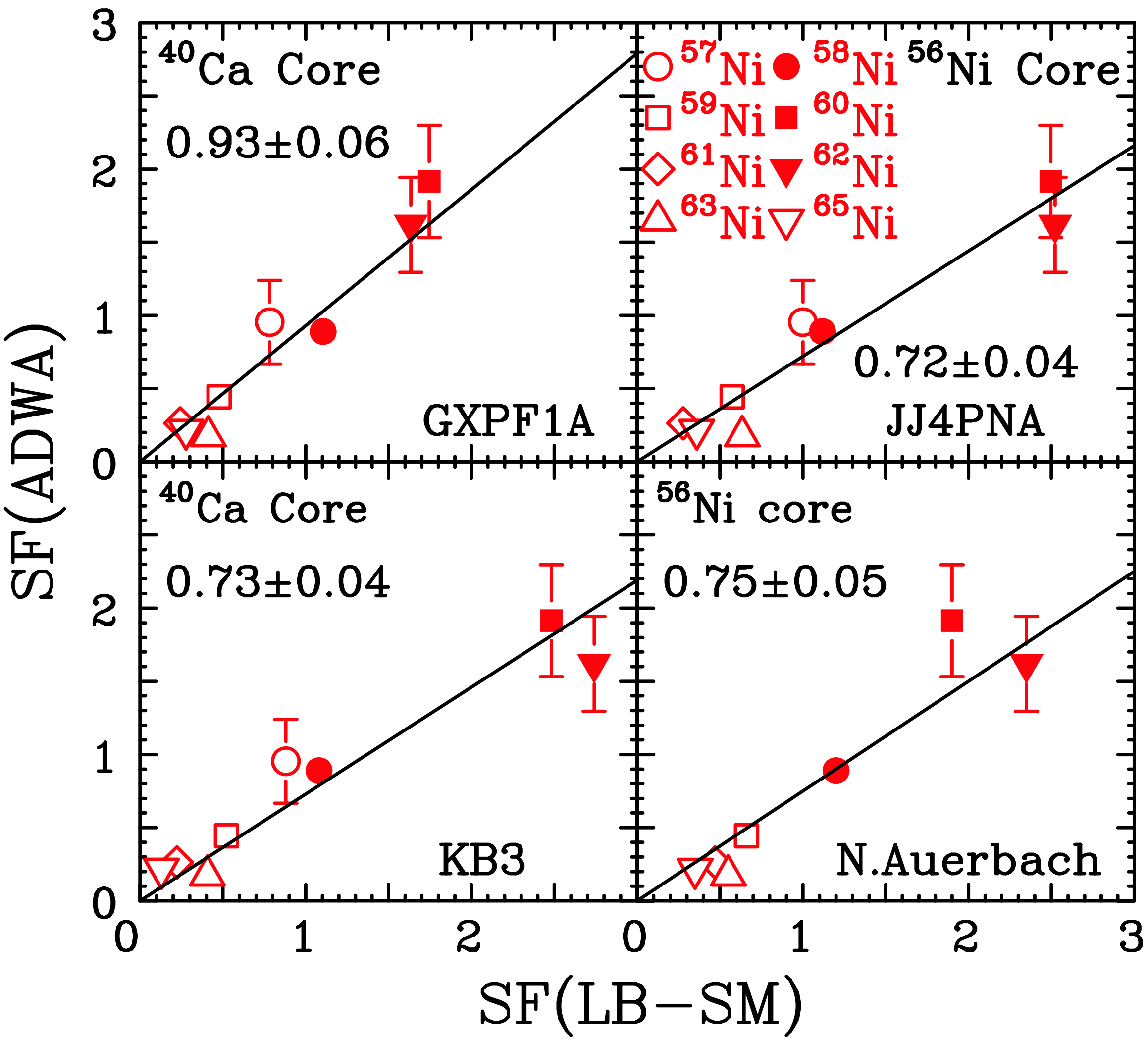

Fig. 3. (Color online) Top panel: Ratios of the experimental SF values divided by values obtained from the shell model calculations with GXPF1A interaction in *pf* model space as a function of the energy levels. Bottom panel: Same as the top panel, but with calculation using JJ4PNA interaction in *gfp* model space. The solid lines indicate perfect agreement and dashed lines represent ±30% of solid line.

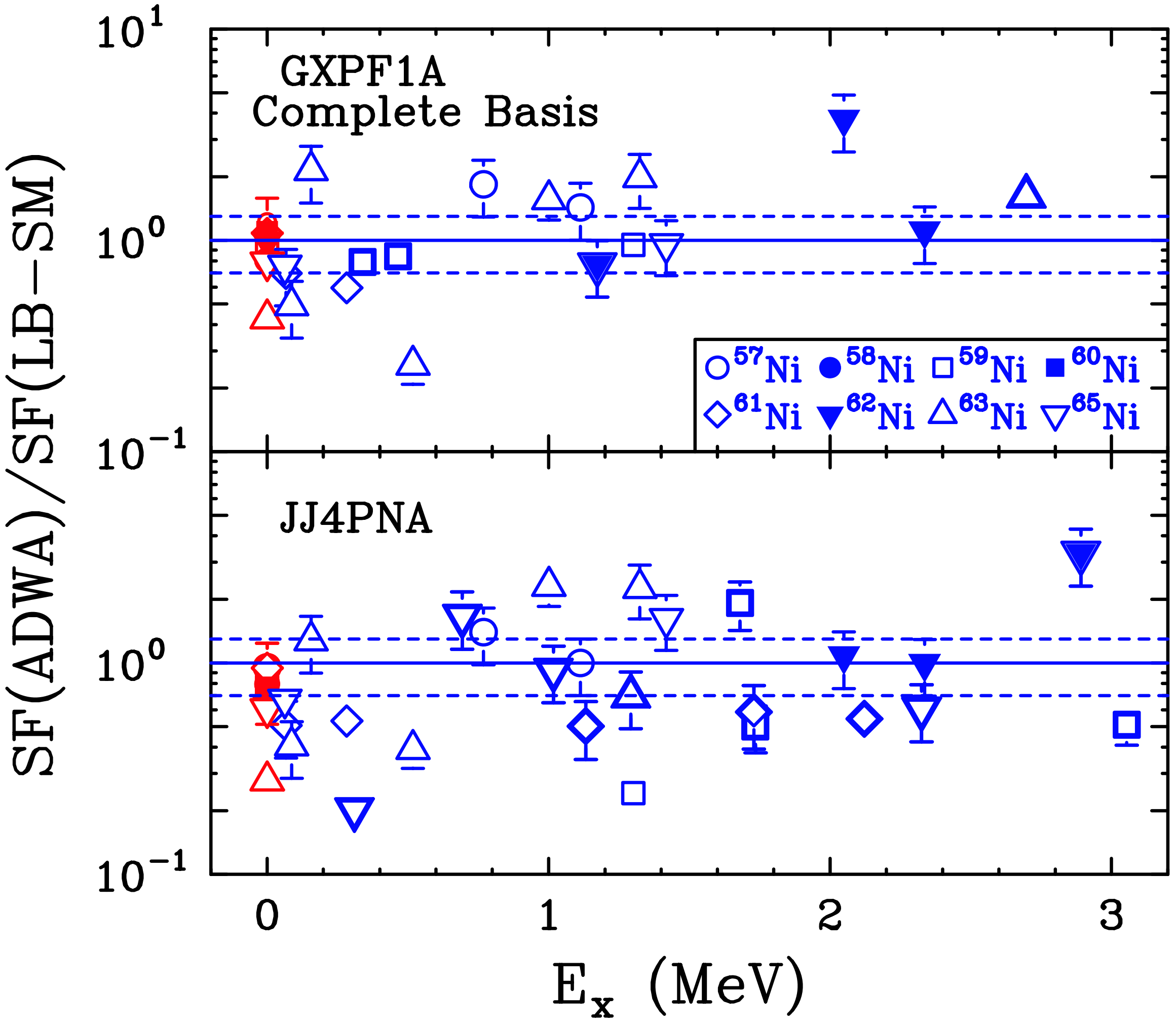

Fig. 4. (Color online) Left panels: Plot of energy levels below 2 MeV for $^{57,61,63}$Ni nuclei with the length of the horizontal bars representing the values of the spectroscopic factors. Right panels: Same as left panels, but for $^{59,62,65}$Ni nuclei. The scale of the SF factor is given in the upper left corner of each panel.

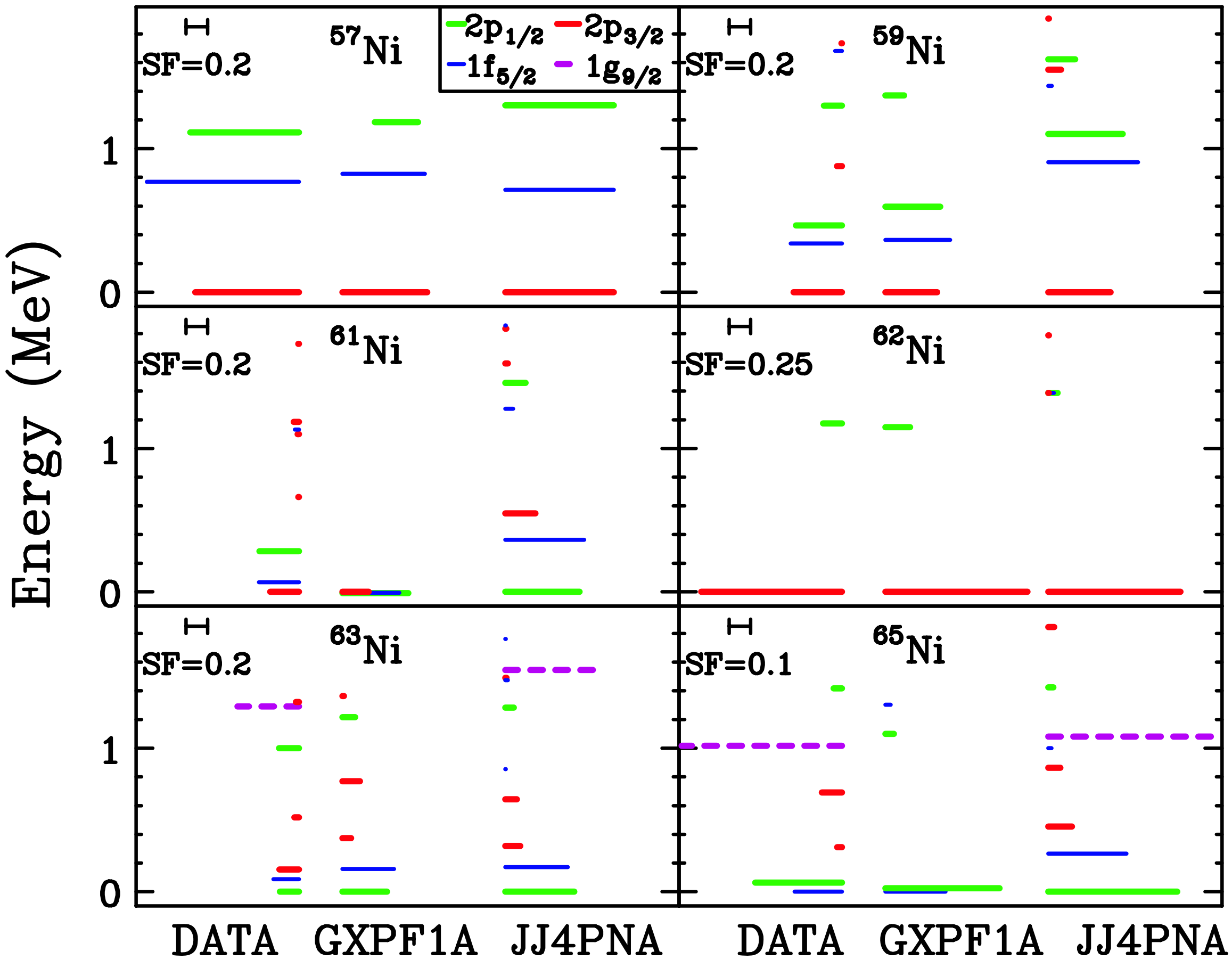

Table I. List of neutron spectroscopic factors for the Ni isotopes. We adopt the energy levels compiled in the data base NUDAT by the National Nuclear Data Center [48]. SF(ADWA) are SF values from the present work and SF(ENDSF) values are obtained from ENDSF data base [48]. Spin value, J, enclosed in ”()” represents state with uncertain J value and the symbol “*” represents doublet state. State with undetermined parity (π) is labeled “N”.

| Nucleus | Ex (MeV) | l | J | π | SF(ADWA) | Error | SF(ENDSF) |
|---|---|---|---|---|---|---|---|
| $^{57}$Ni | 0.000 | 1 | 3/2 | - | 0.954 | ±0.286 | |
| | 0.769 | 3 | 5/2 | - | 1.400 | ±0.420 | |
| | 1.113 | 1 | 1/2 | - | 1.000 | ±0.300 | |
| | | | | | | | |
| $^{58}$Ni | 0.000 | 1 | 0 | + | 0.890 | ±0.087 | 1.250 |
| | | | | | | | |
| $^{59}$Ni | 0.000 | 1 | 3/2 | - | 0.444 | ±0.045 | 0.816 |
| | 0.339 | 3 | 5/2 | - | 0.472 | ±0.059 | 0.677 |
| | 0.465 | 1 | 1/2 | - | 0.424 | ±0.060 | 0.620 |
| | 0.878 | 1 | 3/2 | - | 0.046 | ±0.006 | 0.072 |
| | 1.301 | 1 | 1/2 | - | 0.166 | ±0.031 | 0.286 |
| | 1.680 | 3 | 5/2 | - | 0.062 | ±0.016 | 0.093 |
| | 1.735 | 1 | 3/2 | - | 0.004 | ±0.001 | 0.009 |
| | 1.948 | 3 | 7/2 | - | 0.013 | ±0.007 | 0.037 |
| | 2.415 | 1 | 3/2 | - | 0.013 | ±0.006 | 0.008 |
| | 2.627 | 3 | 7/2 | - | 0.016 | ±0.008 | 0.039 |
| | 2.640 | 1 | (1/2) | - | 0.022 | ±0.007 | |
| | 2.640 | 1 | (3/2) | - | 0.011 | ±0.003 | |
| | 2.681 | 3 | (5/2) | - | 0.019 | ±0.010 | 0.022 |
| | 3.026 | 1 | 1/2 | - | 0.023 | ±0.007 | |
| | 3.026 | 1 | 1/2* | - | 0.009 | ±0.002 | |
| | 3.026 | 3 | (5/2*) | N | 0.016 | ±0.003 | |
| | 3.061 | 4 | 9/2 | + | 0.479 | ±0.096 | |
| | 3.429 | 0 | (1/2) | N | 0.010 | ±0.003 | |
| | 3.452 | 1 | 3/2 | - | 0.022 | ±0.003 | 0.034 |
| | 3.546 | 2 | (5/2) | N | 0.019 | ±0.006 | |
| | 3.652 | 3 | (5/2) | N | 0.018 | ±0.009 | 0.021 |
| | 3.858 | 1 | 3/2 | - | 0.019 | ±0.013 | 0.025 |
| | 4.036 | 1 | (3/2) | - | 0.031 | ±0.016 | 0.012 |
| | 4.506 | 2 | 5/2 | + | 0.175 | ±0.053 | 0.234 |
| | 4.542 | 2 | 5/2 | - | 0.161 | ±0.023 | |
| | 4.709 | 4 | 9/2 | + | 0.049 | ±0.024 | 0.098 |
| | 4.822 | 2 | (5/2) | N | 0.040 | ±0.020 | |
| | 4.939 | 1 | (1/2) | N | 0.054 | ±0.027 | |
| | 5.069 | 1 | 1/2 | - | 0.009 | ±0.003 | 0.017 |
| | 5.149 | 0 | 1/2 | + | 0.065 | ±0.019 | 0.093 |
| | 5.213 | 2 | 5/2 | + | 0.018 | ±0.005 | 0.026 |
| | 5.258 | 2 | (5/2) | N | 0.017 | ±0.009 | |
| | 5.429 | 4 | (9/2) | + | 0.080 | ±0.015 | |
| | 5.458 | 2 | (5/2) | + | 0.151 | ±0.075 | |
| | 5.528 | 0 | 1/2 | + | 0.120 | ±0.060 | |
| | 5.569 | 0 | (1/2) | + | 0.021 | ±0.006 | 0.024 |
| | 5.692 | 0 | 1/2 | + | 0.077 | ±0.023 | 0.126 |
| | 5.894 | 2 | (5/2) | + | 0.014 | ±0.004 | |
| | 6.142 | 1 | 1/2 | - | 0.027 | ±0.005 | |
| | 6.142 | 1 | 3/2 | - | 0.014 | ±0.003 | |
| | 6.206 | 2 | (5/2) | + | 0.023 | ±0.011 | 0.011 |
| | 6.284 | 2 | (5/2) | N | 0.053 | ±0.026 | |
| | 6.380 | 0 | 1/2 | + | 0.039 | ±0.012 | 0.078 |
| | 6.648 | 2 | 3/2 | + | 0.036 | ±0.011 | |
| | 6.648 | 2 | 5/2 | + | 0.024 | ±0.007 | |
| | 7.073 | 0 | 1/2* | + | 0.027 | ±0.008 | 0.029 |
| | 7.073 | 2 | 5/2* | + | 0.007 | ±0.002 | 0.012 |
| | 7.204 | 0 | 1/2* | + | 0.017 | ±0.005 | 0.019 |
| | 7.204 | 2 | 5/2* | + | 0.005 | ±0.001 | 0.012 |
| | 7.302 | 3 | 7/2 | - | 0.011 | ±0.003 | 0.017 |
| | 7.353 | 2 | 5/2 | + | 0.040 | ±0.020 | 0.007 |
| | 7.604 | 2 | 3/2 | + | 0.004 | ±0.001 | |
| | 7.604 | 2 | 5/2 | + | 0.013 | ±0.004 | |
| | | | | | | | |
| $^{60}$Ni | 0.000 | 1 | 0 | + | 1.915 | ±0.383 | 1.640 |
| | | | | | | | |
| $^{61}$Ni | 0.000 | 1 | 3/2 | - | 0.263 | ±0.026 | 0.346 |
| | 0.067 | 3 | 5/2 | - | 0.368 | ±0.110 | 0.507 |
| | 0.283 | 1 | 1/2 | - | 0.363 | ±0.051 | 0.615 |
| | 0.656 | 1 | 1/2 | - | 0.015 | ±0.005 | 0.027 |
| | 1.100 | 1 | 3/2 | - | 0.012 | ±0.004 | 0.027 |
| | 1.132 | 3 | 5/2 | - | 0.036 | ±0.011 | 0.067 |
| | 1.185 | 1 | 3/2 | - | 0.049 | ±0.015 | 0.064 |
| | 1.729 | 1 | 3/2 | - | 0.006 | ±0.002 | 0.011 |
| | 2.122 | 4 | 9/2 | + | 0.499 | ±0.071 | |
| | 2.124 | 1 | 1/2 | - | 0.242 | ±0.034 | |
| | 2.640 | 1 | 1/2 | - | 0.028 | ±0.009 | |
| | 2.640 | 1 | 3/2 | - | 0.014 | ±0.004 | |
| | 2.697 | 2 | 5/2 | + | 0.062 | ±0.019 | 0.087 |
| | 2.765 | 1 | 3/2 | - | 0.008 | ±0.003 | 0.014 |
| | 2.863 | 1 | 1/2 | - | 0.010 | ±0.003 | |
| | 2.863 | 1 | 3/2 | - | 0.005 | ±0.002 | |
| | 3.062 | 0 | 1/2 | + | 0.023 | ±0.007 | |
| | 3.273 | 1 | (3/2) | - | 0.002 | ±0.001 | 0.003 |
| | 3.382 | 1 | 1/2 | - | 0.007 | ±0.002 | |
| | 3.382 | 1 | 3/2 | - | 0.003 | ±0.001 | |
| | 3.506 | 2 | 3/2 | + | 0.158 | ±0.047 | |
| | 3.506 | 2 | 5/2 | + | 0.105 | ±0.031 | |
| | 3.686 | 1 | 1/2 | - | 0.018 | ±0.005 | |
| | 3.686 | 1 | 3/2 | - | 0.009 | ±0.003 | |
| | 4.568 | 2 | (3/2) | + | 0.006 | ±0.002 | |
| | 4.568 | 2 | (5/2) | + | 0.004 | ±0.001 | |
| | 4.600 | 2 | 5/2 | - | 0.004 | ±0.001 | 0.005 |
| | 5.112 | 1 | 1/2 | - | 0.035 | ±0.010 | |
| | 5.112 | 1 | 3/2 | - | 0.018 | ±0.005 | |
| | 5.185 | 0 | 1/2 | + | 0.027 | ±0.008 | 0.051 |
| | 5.309 | 0 | 1/2 | + | 0.012 | ±0.004 | 0.027 |
| | 5.723 | 2 | (3/2) | N | 0.055 | ±0.016 | |
| | 5.723 | 2 | (5/2) | N | 0.036 | ±0.011 | |
| | 5.987 | 0 | 1/2 | + | 0.021 | ±0.006 | |
| | 6.016 | 2 | (3/2) | + | 0.006 | ±0.002 | |
| | 6.016 | 2 | (5/2) | + | 0.004 | ±0.001 | |
| | 6.346 | 2 | 3/2 | + | 0.019 | ±0.006 | |
| | 6.346 | 2 | 5/2 | + | 0.013 | ±0.004 | |
| | 6.371 | 2 | 3/2 | + | 0.008 | ±0.002 | |
| | 6.371 | 2 | 5/2 | + | 0.006 | ±0.002 | |
| | 6.609 | 2 | 3/2 | + | 0.005 | ±0.002 | |
| | 6.609 | 2 | 5/2 | + | 0.004 | ±0.001 | |
| | | | | | | | |
| $^{62}$Ni | 0.000 | 1 | 0 | + | 1.619 | ±0.324 | |
| | 1.173 | 1 | 2 | + | 0.218 | ±0.065 | |
| | 2.049 | 1 | 0 | + | 0.280 | ±0.084 | |
| | 2.336 | 3 | 4 | + | 0.274 | ±0.082 | |
| | 2.891 | 1 | 0 | + | 0.505 | ±0.152 | |
| | 3.059 | 3 | 2 | + | 0.233 | ±0.070 | |
| | 3.158 | 1 | 2 | + | 0.052 | ±0.016 | |
| | 3.262 | 3 | (2) | + | 1.119 | ±0.336 | |
| | 3.370 | 1 | 1 | + | 0.295 | ±0.089 | |
| | 3.370 | 1 | 2 | + | 0.177 | ±0.053 | |
| | 3.519 | 1 | 2 | + | 0.248 | ±0.074 | |
| | 3.757 | 4 | 3 | - | 0.361 | ±0.108 | |
| | 3.849 | 1 | 0 | + | 1.028 | ±0.309 | |
| | 3.849 | 1 | 1 | + | 0.343 | ±0.103 | |
| | 3.849 | 1 | 2 | + | 0.206 | ±0.062 | |
| | 4.393 | 3 | (2) | N | 0.144 | ±0.043 | |
| | 4.503 | 4 | (3) | - | 0.264 | ±0.079 | |
| | 4.720 | 4 | (3) | - | 0.791 | ±0.237 | |
| | 4.863 | 4 | 5 | - | 1.079 | ±0.324 | |
| | 4.863 | 4 | 6 | - | 0.913 | ±0.274 | |
| | 5.331 | 2 | (3) | - | 0.163 | ±0.049 | |
| | 5.545 | 4 | 3 | - | 0.653 | ±0.196 | |
| | 5.545 | 4 | 4 | - | 0.508 | ±0.152 | |
| | 5.545 | 4 | 5 | - | 0.416 | ±0.125 | |
| | 5.545 | 4 | 6 | - | 0.352 | ±0.106 | |
| | 5.628 | 2 | 3 | - | 0.024 | ±0.007 | |
| | 6.103 | 2 | 1 | - | 0.451 | ±0.135 | |
| | 6.103 | 2 | 2 | - | 0.270 | ±0.081 | |
| | 6.103 | 2 | 3 | - | 0.193 | ±0.058 | |
| | 6.103 | 2 | 4 | - | 0.150 | ±0.045 | |
| | 6.540 | 2 | 1 | - | 0.350 | ±0.105 | |
| | 6.540 | 2 | 2 | - | 0.210 | ±0.063 | |
| | | | | | | | |
| $^{63}$Ni | 0.000 | 1 | 1/2 | - | 0.176 | ±0.025 | 0.370 |
| | 0.087 | 3 | 5/2 | - | 0.234 | ±0.070 | 0.563 |
| | 0.156 | 1 | 3/2 | - | 0.177 | ±0.053 | 0.275 |
| | 0.518 | 1 | 3/2 | - | 0.042 | ±0.008 | 0.080 |
| | 1.001 | 1 | 1/2 | - | 0.184 | ±0.037 | 0.330 |
| | 1.292 | 4 | (9/2) | + | 0.565 | ±0.169 | 0.750 |
| | 1.324 | 1 | 3/2 | - | 0.028 | ±0.008 | 0.063 |
| | 2.297 | 2 | 5/2 | + | 0.189 | ±0.027 | 0.142 |
| | 2.697 | 1 | 1/2 | - | 0.023 | ±0.003 | 0.045 |
| | 2.953 | 0 | 1/2 | + | 0.128 | ±0.038 | 0.190 |
| | 3.104 | 2 | 3/2 | + | 0.016 | ±0.005 | |
| | 3.104 | 2 | 5/2 | + | 0.011 | ±0.003 | |
| | 3.283 | 2 | (5/2) | N | 0.041 | ±0.012 | 0.053 |
| | 3.292 | 2 | 5/2 | + | 0.037 | ±0.011 | |
| | 3.740 | 2 | (3/2) | N | 0.030 | ±0.009 | 0.040 |
| | 3.951 | 2 | 5/2 | + | 0.074 | ±0.022 | 0.100 |
| | 4.387 | 2 | 5/2 | + | 0.038 | ±0.011 | 0.062 |
| | 4.622 | 2 | 3/2 | + | 0.053 | ±0.016 | |
| | 4.622 | 2 | 5/2 | + | 0.036 | ±0.005 | |
| | 5.060 | 2 | (3/2) | + | 0.009 | ±0.003 | |
| | 5.060 | 2 | (5/2) | + | 0.006 | ±0.002 | |
| | | | | | | | |
| $^{65}$Ni | 0.000 | 3 | 5/2 | - | 0.218 | ±0.031 | 0.338 |
| | 0.063 | 1 | 1/2 | - | 0.399 | ±0.056 | 0.620 |
| | 0.310 | 1 | 3/2 | - | 0.022 | ±0.003 | 0.035 |
| | 0.693 | 1 | 3/2 | - | 0.093 | ±0.028 | 0.235 |
| | 1.017 | 4 | 9/2 | + | 0.738 | ±0.221 | 0.085 |
| | 1.418 | 1 | 1/2 | - | 0.038 | ±0.011 | 0.257 |
| | 1.920 | 2 | 5/2 | + | 0.173 | ±0.052 | |
| | 2.163 | 1 | (1/2) | N | 0.031 | ±0.009 | |
| | 2.325 | 3 | (5/2*) | N | 0.030 | ±0.009 | |
| | 2.325 | 4 | (9/2*) | N | 0.050 | ±0.015 | |
| | 2.336 | 3 | (5/2) | N | 0.085 | ±0.025 | |
| | 2.336 | 3 | (7/2) | N | 0.063 | ±0.019 | 0.003 |
| | 2.712 | 2 | 3/2 | + | 0.003 | ±0.001 | |
| | 3.044 | 1 | (1/2) | N | 0.022 | ±0.007 | |
| | 3.044 | 1 | (3/2) | N | 0.011 | ±0.003 | |
| | 3.411 | 2 | (3/2) | + | 0.130 | ±0.039 | |
| | 3.411 | 2 | (5/2) | + | 0.087 | ±0.026 | |
| | 3.463 | 2 | (3/2) | N | 0.008 | ±0.002 | |
| | 3.463 | 2 | (5/2) | N | 0.005 | ±0.002 | 0.082 |
| | 3.563 | 2 | 5/2 | + | 0.065 | ±0.013 | 0.042 |
| | 3.743 | 2 | 5/2 | + | 0.031 | ±0.009 | 0.068 |
| | 3.907 | 2 | 5/2 | + | 0.058 | ±0.018 | |
| | 4.391 | 2 | 3/2 | + | 0.057 | ±0.017 | |
| | 4.391 | 2 | 5/2 | + | 0.038 | ±0.011 | |

Table II. Comparison of experimental and large-basis shell-model energy levels and spectroscopic factors for Ni isotopes.

| | | | | Ex (MeV) | | | SF | | | |
|---|---|---|---|---|---|---|---|---|---|---|
| Nucleus | l | J | P | NUDAT | GXPF1A | JJ4PNA | ADWA | Error | GXPF1A | JJ4PNA |
| $^{57}$Ni | 1 | 3/2 | - | 0 | 0 | 0 | 0.954 | ±0.286 | 0.783 | 1.000 |
| | 3 | 5/2 | - | 0.769 | 0.825 | 0.714 | 1.400 | ±0.42 | 0.76 | 1.000 |
| | 1 | 1/2 | - | 1.113 | 1.184 | 1.302 | 1.000 | ±0.3 | 0.698 | 1.000 |
| $^{58}$Ni | 1 | 0 | + | 0 | 0 | 0 | 0.890 | ±0.087 | 1.105 | 1.118 |
| $^{59}$Ni | 1 | 3/2 | - | 0 | 0 | 0 | 0.444 | ±0.045 | 0.477 | 0.574 |
| | 3 | 5/2 | - | 0.339 | 0.364 | | 0.472 | ±0.059 | 0.597 | |
| | 1 | 1/2 | - | 0.465 | 0.595 | | 0.424 | ±0.06 | 0.504 | |
| | 1 | 1/2 | - | 1.301 | 1.371 | 1.103 | 0.166 | ±0.031 | 0.175 | 0.685 |
| | 3 | 5/2 | - | 1.68 | | 1.439 | 0.062 | ±0.016 | | 0.032 |
| | 1 | 3/2 | - | 1.735 | | 1.906 | 0.004 | ±0.001 | | 0.008 |
| | 4 | 9/2 | + | 3.061 | | 3.454 | 0.479 | ±0.096 | | 0.938 |
| | 4 | 9/2 | + | 4.709 | | 4.540 | 0.049 | ±0.024 | | 0.007 |
| | 4 | 9/2 | + | 5.429 | | 5.418 | 0.080 | ±0.015 | | 0.028 |
| $^{60}$Ni | 1 | 0 | + | 0 | 0 | 0 | 1.915 | ±0.383 | 1.746 | 2.496 |
| $^{61}$Ni | 1 | 3/2 | - | 0 | 0 | 0.547 | 0.263 | ±0.026 | 0.244 | 0.278 |
| | 3 | 5/2 | - | 0.067 | -0.006 | 0.364 | 0.368 | ±0.11 | 0.527 | 0.727 |
| | 1 | 1/2 | - | 0.283 | -0.008 | 0 | 0.363 | ±0.051 | 0.609 | 0.683 |
| | 1 | 1/2 | - | 0.656 | | 1.457 | 0.015 | ±0.005 | | 0.188 |
| | 3 | 5/2 | - | 1.132 | | 1.277 | 0.036 | ±0.011 | | 0.072 |
| | 1 | 3/2 | - | 1.729 | | 1.835 | 0.006 | ±0.002 | | 0.010 |
| | 4 | 9/2 | + | 2.122 | | 2.516 | 0.499 | ±0.071 | | 0.917 |
| | 1 | 1/2 | - | 2.124 | | 2.280 | 0.242 | ±0.034 | | 0.007 |
| | 1 | 3/2 | - | 3.686 | | 3.669 | 0.009 | ±0.003 | | 0.001 |
| $^{62}$Ni | 1 | 0 | + | 0 | 0 | 0 | 1.619 | ±0.324 | 1.635 | 2.522 |
| | 1 | 2 | + | 1.173 | 1.148 | | 0.218 | ±0.065 | 0.284 | |
| | 1 | 0 | + | 2.049 | 2.188 | 2.263 | 0.280 | ±0.084 | 0.075 | 0.259 |
| | 3 | 4 | + | 2.336 | 2.256 | 2.317 | 0.274 | ±0.082 | 0.247 | 0.275 |
| | 1 | 0 | + | 2.891 | | 2.740 | 0.505 | ±0.152 | | 0.153 |
| $^{63}$Ni | 1 | 1/2 | - | 0 | 0 | 0 | 0.176 | ±0.025 | 0.412 | 0.634 |
| | 3 | 5/2 | - | 0.087 | 0.158 | 0.171 | 0.234 | ±0.07 | 0.476 | 0.576 |
| | 1 | 3/2 | - | 0.156 | 0.373 | 0.319 | 0.177 | ±0.053 | 0.083 | 0.138 |
| | 1 | 3/2 | - | 0.518 | 0.77 | 0.643 | 0.042 | ±0.008 | 0.163 | 0.107 |
| | 1 | 1/2 | - | 1.001 | 1.216 | 1.282 | 0.184 | ±0.037 | 0.118 | 0.079 |
| | 4 | 9/2 | + | 1.292 | | 1.546 | 0.565 | ±0.169 | | 0.811 |
| | 1 | 3/2 | - | 1.324 | 1.363 | 1.491 | 0.028 | ±0.008 | 0.014 | 0.012 |
| | 1 | 1/2 | - | 2.697 | 2.79 | | 0.023 | ±0.003 | 0.014 | |
| $^{65}$Ni | 3 | 5/2 | - | 0 | 0 | 0.264 | 0.218 | ±0.031 | 0.277 | 0.360 |
| | 1 | 1/2 | - | 0.063 | 0.025 | 0 | 0.399 | ±0.056 | 0.526 | 0.594 |
| | 1 | 3/2 | - | 0.310 | | 0.453 | 0.022 | ±0.003 | | 0.109 |
| | 1 | 3/2 | - | 0.693 | | 0.864 | 0.093 | ±0.028 | | 0.056 |
| | 4 | 9/2 | + | 1.017 | | 1.082 | 0.738 | ±0.221 | | 0.797 |
| | 1 | 1/2 | - | 1.418 | 1.100 | 1.425 | 0.038 | ±0.011 | 0.040 | 0.024 |
| | 4 | 9/2 | + | 2.325 | | 2.474 | 0.05 | ±0.015 | | 0.083 |